\documentclass[twocolumn]{aastex63}
\usepackage{epsfig}
\usepackage{enumitem}

\begin{document}

\title{The evolution effects of radius and moment of inertia for rapidly rotating neutron stars}

%\title{The evolution effects of radius and moment of inertia for rapidly rotating neutron stars}

\author{Lin Lan}
\affiliation{Department of Astronomy, Beijing Normal University, Beijing 100875, China; gaohe@bnu.edu.cn}

\author{He Gao}
\affiliation{Department of Astronomy, Beijing Normal University, Beijing 100875, China; gaohe@bnu.edu.cn}

\author{Shunke Ai}
\affiliation{Department of Physics and Astronomy, University of Nevada Las Vegas, Las Vegas, NV 89154, USA}

\author{Shao-Ze Li}
\affiliation{Department of Astronomy, Beijing Normal University, Beijing 100875, China; gaohe@bnu.edu.cn}
\begin{abstract}
A newly born millisecond magnetar is thought to be the central engine of some gamma-ray bursts (GRBs), especially those that present long-lasting X-ray plateau emissions. By solving the field equations, we find that when the rotational speed of the magnetar is approaching the breakup limit, its radius $R$ and moment of inertia $I$ would undergo an obvious evolution as the magnetar spins down. Meanwhile, the values of $R$ and $I$ would sensitively depend on the adoption of neutron star (NS) equation of state (EoS) and the NS baryonic mass. With different EoSs and baryonic masses considered, the magnetic dipole radiation luminosity ($L_{\rm dip}$) could be variant within one to two orders of magnitude. We thus suggest that when using the X-ray plateau data of GRBs to diagnose the properties of the nascent NSs, EoS and NS mass information should be invoked as simultaneously constrained parameters. On the other hand, due to the evolution of $R$ and $I$, the temporal behavior of $L_{\rm dip}$ would become more complicated. For instance, if the spin-down process is dominated by gravitational wave emission due to the NS asymmetry caused by magnetic field distortion ($\epsilon\propto B_{p}^{2}$), the segment $L_{\rm dip}\propto t^{0}$ could be followed by $L_{\rm dip}\propto t^{-\gamma}$ with $\gamma$ larger than 3. This case could naturally interpret the so-called internal X-ray plateau feature shown in some GRB afterglows, which means the sharp decay following the plateau is unnecessarily corresponding to the NS collapsing. This may explain why some internal X-ray plateaus are followed by late time central engine activity, manifested through flares and second shallow plateaus.

\end{abstract}

%%%%%%%%%%%%%%%%%%%%%%%%%%%%%%%%%%%%%%%%%%%%%%%%%%%%%%%%%%%%%%%%%%%%%%%%%%%%%%%%%%%%%%%%%%%%%%%%%%%%%%%%%%%%%%%
%%%%%%%%%%%%%%%%%%%%%%%%%%%%%%%%%%%%%%%%%%%%%%%%%%%%%%%%%%%%%%%%%%%%%%%%%%%%%%%%%%%%%%%%%%%%%%%%%%%%%%%%%%%%%%%

\section{INTRODUCTION}

With the successful launch of the \emph{Neil Gehrels Swift Observatory} \citep{Gehrels04}, early-time
afterglows for gamma-ray bursts (GRBs) have been revealed. They are important for unveiling the open
questions for GRB physics, such as their progenitors and central engines \citep[][for a review]{zhang18}.
Observationally, a good fraction of the X-ray light curves of (long and short) GRBs were discovered
to show a long-lasting plateau feature. Such a feature seems to be consistent with the prediction
of millisecond magnetar (namely rapidly spinning, strongly magnetized neutron star) engine, which
is thought to be formed from a violent event such as the massive star collapse (for long GRBs) or
double neutron star merger (for short GRBs) \citep{Paczynski86,Eichler89,Usov92,Woosley93,Thompson94,Dai98a,Dai98b,MacFadyen99,Zhang01,Metzger08,Zhang11a}.

The energy reservoir of a newly born millisecond magnetar is the total rotational energy, which could be
estimated as
\begin{eqnarray}
E_{\rm rot} = \frac{1}{2} I \Omega^{2}
\label{Erot}
\end{eqnarray}
where $I$ is the moment of inertia and $\Omega=2\pi/P$ is the angular frequency of the nascent neutron star (NS).
In principle, during the spin-down process, the nascent magnetar loses its rotational energy via both magnetic dipole
radiation and gravitational wave (GW) emission, so that the spin-down law can be written as \citep{Shapiro83,Zhang01}
\begin{eqnarray}
\dot{E}&=-L_{\rm dip}-L_{\rm GW}&=-\frac{B_p^2R^6\Omega^4}{6c^3}-\frac{32GI^2\epsilon^2\Omega^6}{5c^5}.
\label{dotE}
\end{eqnarray}
$B_p$ is the dipolar field strength at the magnetic poles on the NS surface. $R$ and $\epsilon$ are the radius
and ellipticity of the NS, respectively. Depending on the NS properties, such as the values of $B_p$ and $\epsilon$,
the spin-down process will be dominated by one or the other loss term.

The Poynting flux generated by the EM dipolar emission could undergo the magnetic energy dissipation processes with high
efficiency \citep{Zhang11b} and power the X-ray plateau feature in GRB afterglows \citep{Rowlinson10,Rowlinson13,Gompertz13,Gompertz14,Lu14,Lu15}.
In principle, the physical parameters of the newly born magnetar could be estimated by fitting the observed GRB
X-ray plateau data. On the other hand, one can also infer {the dominant energy loss term} (EM or GW radiation) by
analyzing the braking index of the magnetar, which can be obtained by fitting the
decay slopes of the X-ray plateau and its follow-up segment  \citep{Zhang01,Lasky16,Lasky17,Lu18,Lu19}. Studying the population of braking indices is another feasible way to look at this \citep{Sarin20a}.
Some GRBs were discovered to show an extended plateau, followed by a sharp decay with a decay index larger than 3 \citep[called internal plateau;][]{Rowlinson10,Rowlinson13,Lu15,Sarin20a},
which is usually interpreted as the magnetar collapses into a BH \footnote{Alternatively, it has been proposed that the internal plateau might be an imprint of the r-process heating on fall-back accretion \citep{Desai19}, or be the high-latitude emission signature from a structured jet  \citep{Ascenzi20}.}. For these cases, one can make estimation for the dipolar
magnetic field strength $B_p$ and the initial spin period $P_0$ of the magnetar from the observed X-ray plateau luminosity and its
ending time \citep{Rowlinson10,Rowlinson13,Gompertz13,Gompertz14,Lu15,Gao16,Sarin20a}.

In the previous studies, $R$ and $I$ of magnetars are usually assumed to be constant, and are usually assigned to some so-called fiducial
values, such as $R=10^6~\rm cm$ and $I=1.5\times10^{45}~\rm g~cm^2$. However, as the GRB central engine, the magnetars are supposed to be
rapidly rotating (especially for short GRBs where the magnetars are formed from merger), in which case its radius and moment of inertia
should be related to the rotational speed \citep{lattimer12}. Nevertheless, the specific values of $R$ and $I$ should also depend on the
NS equation of state (EoS) and the NS mass.

In this paper, we intend to study in detail how $R$ and $I$ evolve as the magnetar spins down, how these evolution effects alter the
magnetic dipole radiation behavior, and to what extent the NS EoS and NS mass affect the value of dipole radiation luminosity.

\section{$R$ and $I$ evolution}

For the purpose of this work, we first need to know how the radius $R$ and moment of inertia $I$ for a given fast
rotating neutron star change with its rotational speed. Obviously, it depends on the EoS of neutron star. Here we
use the numerical methods to treat the equilibrium equations for a stationary, axially symmetric, rigid rotating NS,
within a fully general relativistic framework. In this case, the spacetime metric
can be written as
\begin{eqnarray}
ds^2 &=& -e^{2\nu}\, dt^2 + r^2 \sin^2\theta B^2 e^{-2\nu}
(d\phi - \omega\, dt)^2
\nonumber \\ & & \mbox{}
+ e^{2\alpha} (dr^2 + r^2\, d\theta^2),
\end{eqnarray}
where the potentials $\nu$, $B$, $\omega$, and $\alpha$ depend only on $r$ and $\theta$. We have the
following asymptotic decay as \citep{Butterworth76}
\begin{eqnarray}
\nu &=& -\frac{M}{r} + \frac{B_0M}{3r^3} + \nu_2 P_2(\cos\theta)+\mathcal{O}\left(\frac{1}{r^{4}}\right), \nonumber \\
B &=& 1 + \frac{B_0}{r^2} +\mathcal{O}\left(\frac{1}{r^{4}}\right) , \nonumber \\
\omega &=&\frac{2I\Omega}{r^3}+\mathcal{O}\left(\frac{1}{r^{4}}\right) ,
\end{eqnarray}
where $M$ is the NS mass and $\Omega$ is the angular frequency. $B_0$ and $\nu_2$ are real constant.
When describing the interior of the NS as a perfect fluid, its energy-momentum tensor becomes
\begin{eqnarray}
T^{\mu\nu} = (\rho + p)u^{\mu}u^{\nu} + p g^{\mu\nu},
\end{eqnarray}
where $\rho$ presents the energy density, $p$ denotes the pressure, and $u^{\mu}$ is the
$4$-velocity.

\begin{figure}[t]
%\centering
\includegraphics	[angle=0,scale=0.6]	{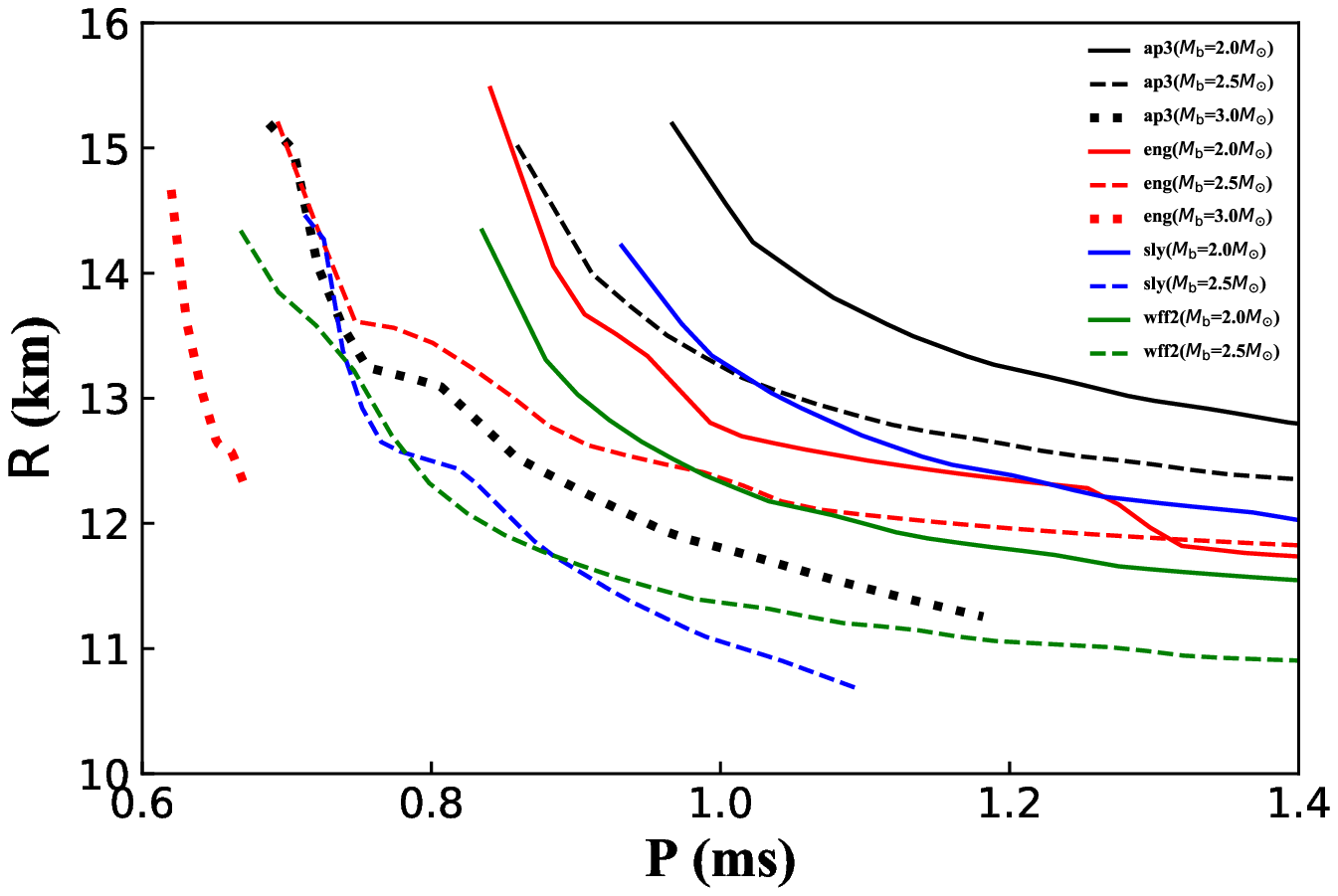}
\includegraphics	[angle=0,scale=0.6]	{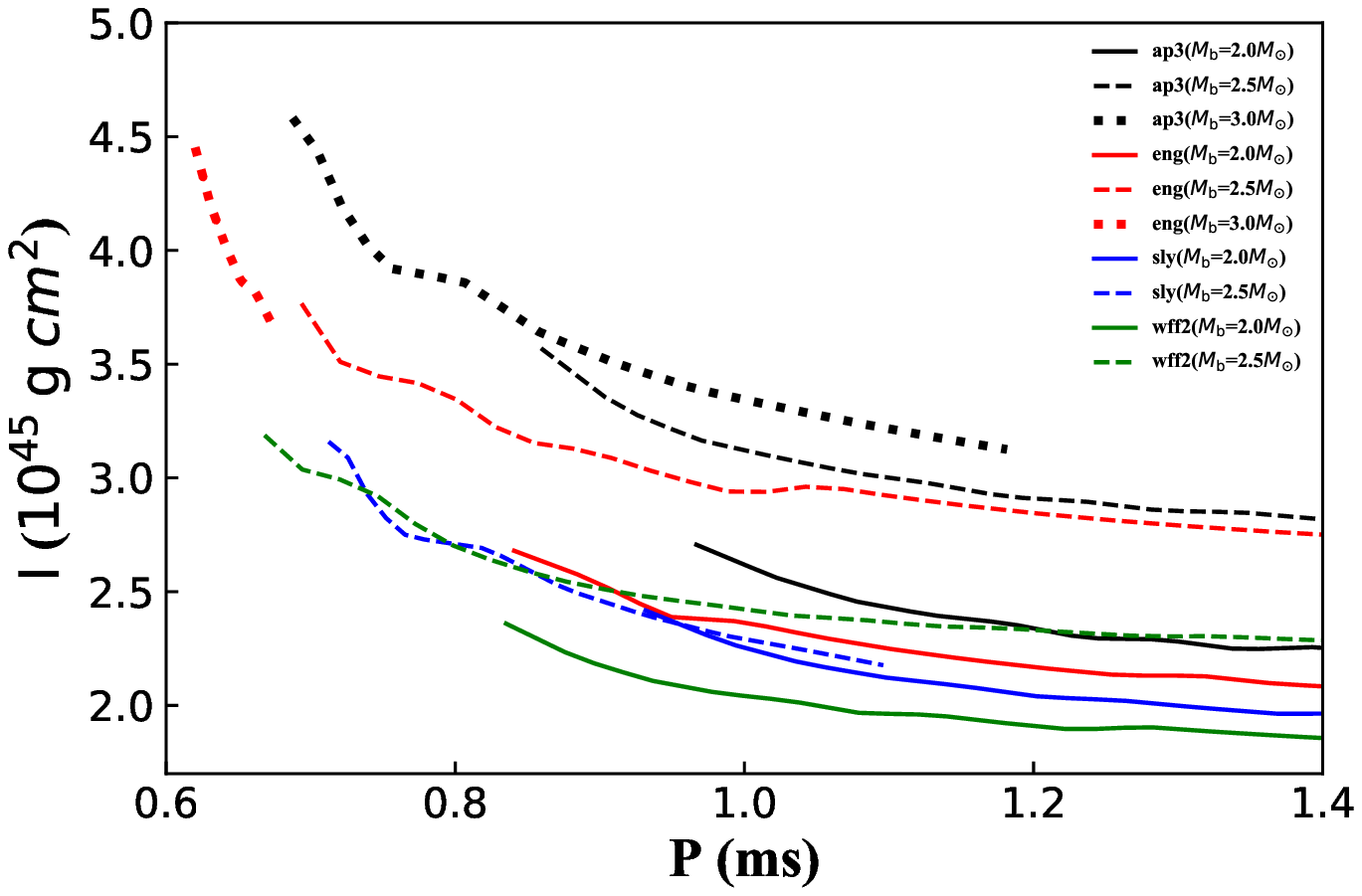}\\
\caption{Evolutionary behavior of the neutron star radius ($R$) and moment of inertia ($I$) for different EoSs and baryonic mass.}
\nonumber
\label{fig:R-I}
\end{figure}

For a selected sample of EoSs [SLy \citep{Douchin01}, ENG \citep{Engvik96}, AP3 \citep{Akmal97}, WFF2 \citep{Wiringa88}] within
a range of maximum mass ($2.05M_{\odot}<M_{\rm TOV}<2.39M_{\odot}$), we use the public code \texttt{RNS} \citep{Stergioulas95} to
solve the above field equations for $R-P$ and $I-P$ relationships, where $P$ is the NS spinning period. For completeness, for a
given EoS, we also consider the situations when the NSs have different baryonic mass $M_b$. The results are presented in Figure \ref{fig:R-I}.

For a given EoS, when the NS rotational speed is approaching the breakup limit, $R$ and $I$ approach to constant. As the NS
spins down, both $R$ and $I$ would go through a rapid descent phase. When the spin is slow enough, $R$ and $I$ again tend to
be constant. For a given rotational speed, greater baryonic mass of the NS would lead to smaller $R$ and greater $I$.
For our selected EoSs, when the baryonic mass or rotational speed of a NS undergoes changes, the radius of the NS can vary from 10 km
to 15 km, and the moment of inertia of the NS can vary from $1.5\times10^{45}~\rm g~cm^{2}$ to $4.5\times10^{45}~\rm g~cm^{2}$.

\section{Numerical Results}

With the solutions of $R$ and $I$ for various rotational speed, for a NS with given parameters (e.g., given the initial values of $B_p$, $\epsilon$, $M_b$ and $P_0$), we can numerically solve its spin-down evolution history, and calculate the evolution history of magnetic dipole radiation luminosity. The result is obviously dependent on the NS EoS and different baryonic mass. For each of our selected EoSs and baryonic mass, we calculate the spin-down solutions for a large sample of NSs with various parameter combinations, where the NS spin-down process is dominated either by EM radiation or GW radiation. Here we assume that the magnetic dipole moment $\mu\equiv B_p R^3$ is conserved.

\begin{figure}
%\centering
\includegraphics	[angle=0,scale=0.6]	{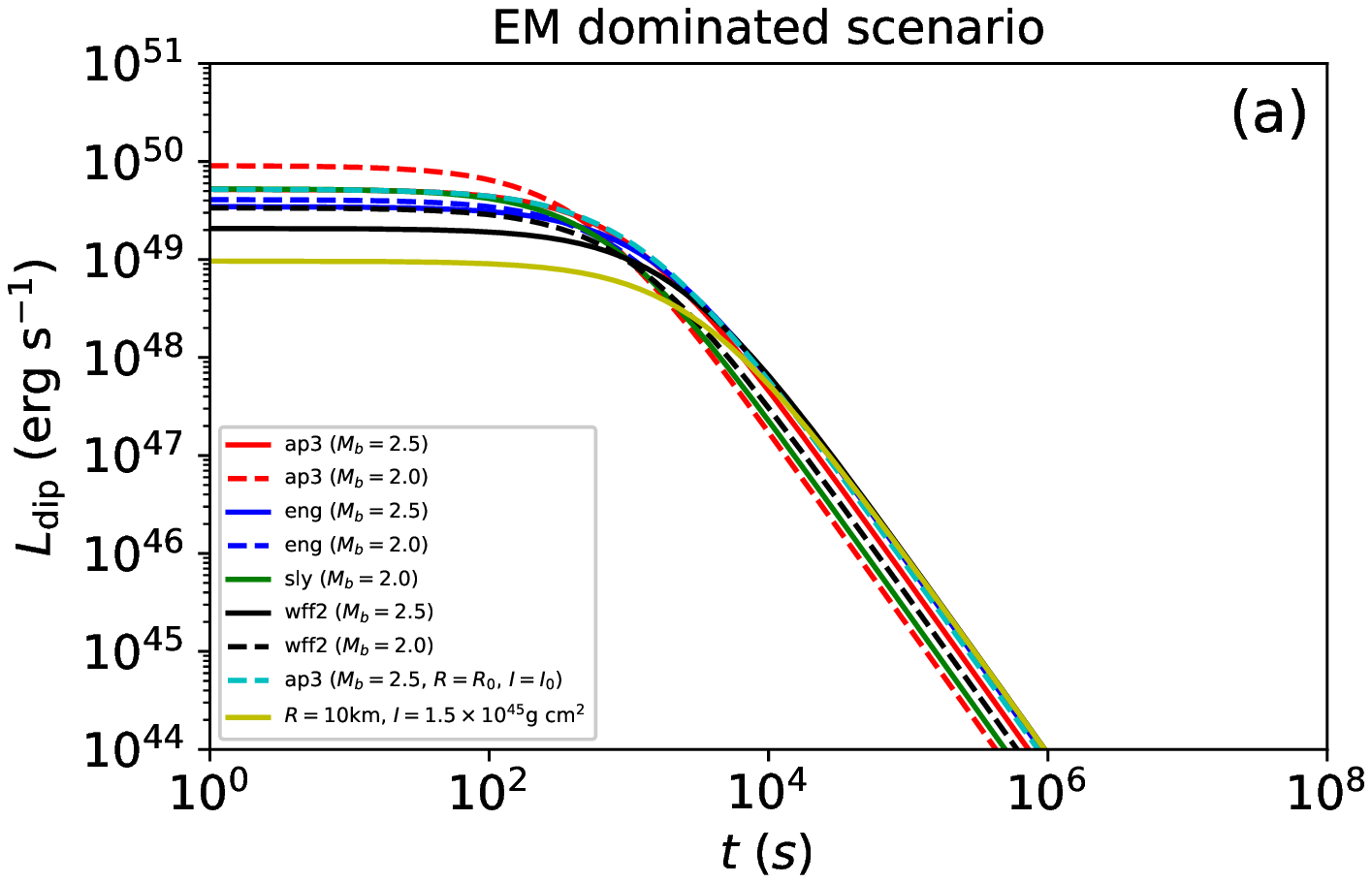}
\includegraphics	[angle=0,scale=0.6]	{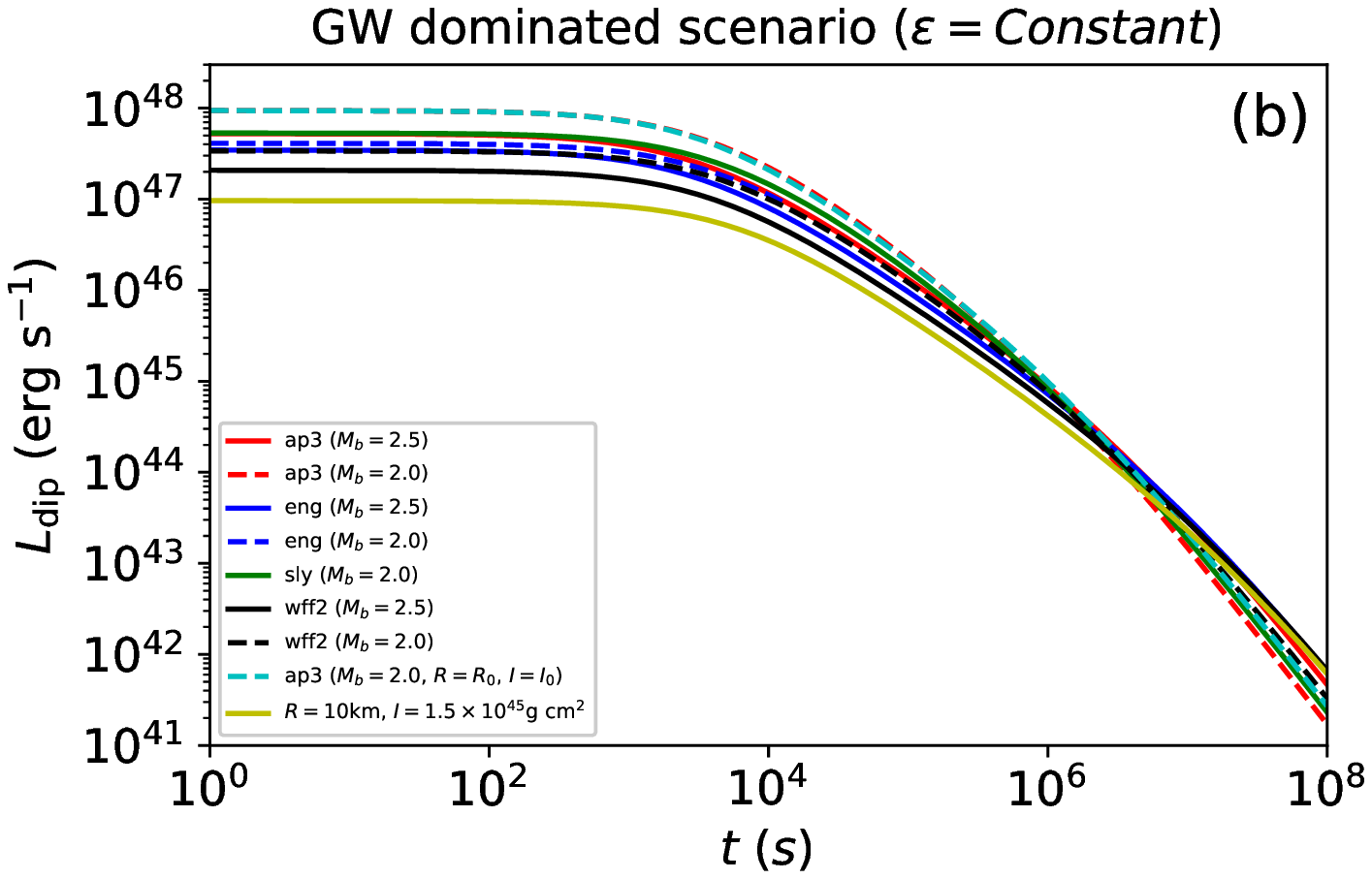}
\includegraphics	[angle=0,scale=0.6]	{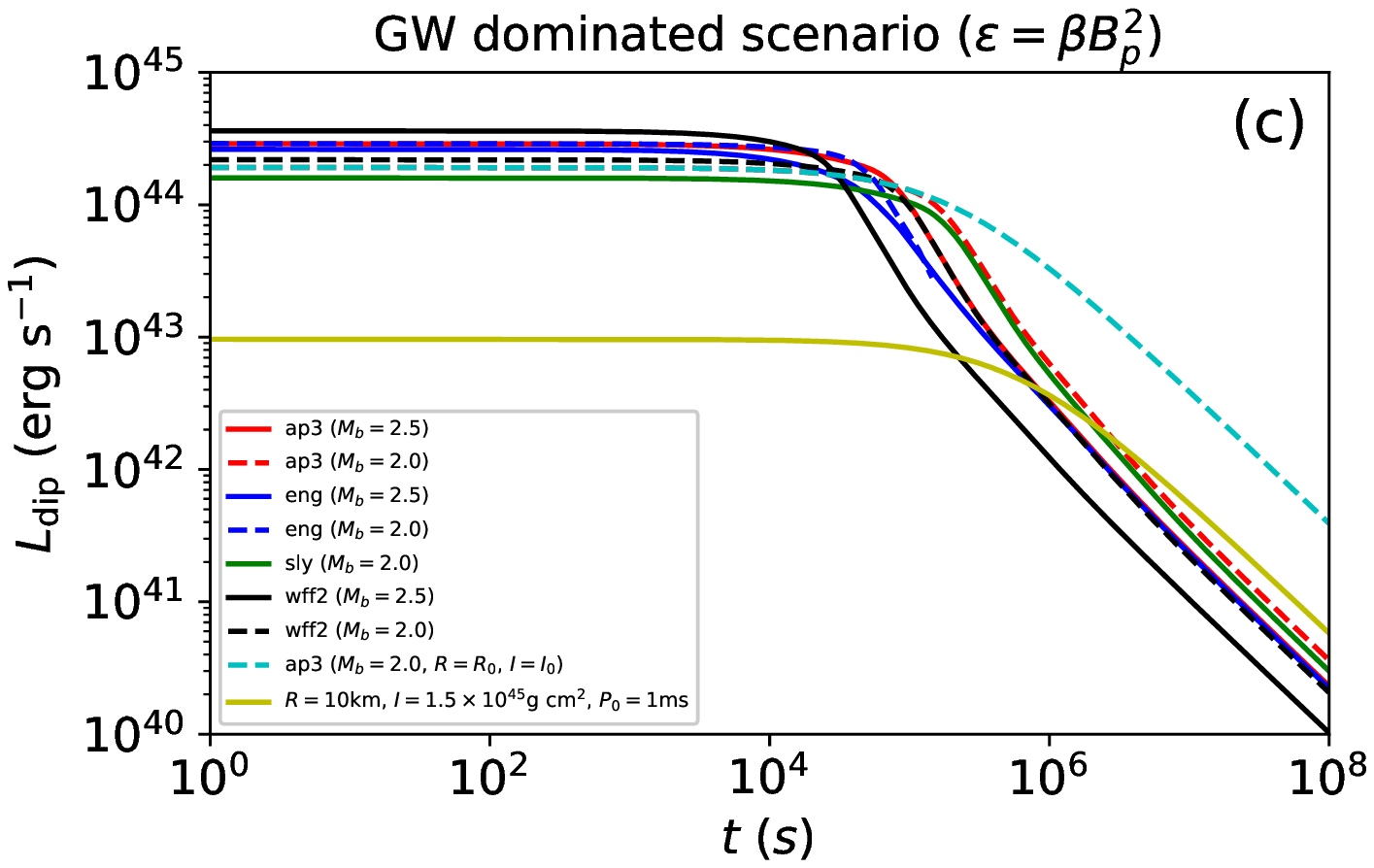}\\
\caption{Numerical results for dipole radiation luminosity by considering the $R$ and $I$ evolution effects in choice of different NS EoSs and baryonic mass of NS.(a): EM dominated scenario. We set $B_{p,0} = 10^{15} G$ and $P_0 = 1 {\rm ms}$ for initial condition, and $\epsilon = 10^{-7}$; (b): GW dominated scenario with a constant ellipticity. We set $B_{p,0} = 10^{14} G$ and $P_0 = 1 {\rm ms}$ for initial condition, and $\epsilon = 10^{-3}$; (c): GW dominated scenario with $\epsilon\propto B_{p}^{2}$. We set the initial condition as $B_{p,0} = 10^{12} G$, $\epsilon_0 = 10^{-4}$. Here we use the Kepler period for each adopted $M_b$ and EoS as the initial spinning period $P_0$.}
\nonumber
\label{fig:Ldip}
\end{figure}

Firstly, for EM radiation dominated case, we set initial value $B_{p,0}=10^{15}$ G, $P_0 =1 {\rm ms}$  and $\epsilon=10^{-7}$.
We test two values for $M_b$ (e.g. $2.0M_{\odot}$ and $2.5M_{\odot}$), with which the NS can be supported by most of the adopted EoSs.
%We test three values (e.g., 0.8 ms, 1 ms and 2 ms) for $P_0$, and $M_b$ is adopted from 2 $M_\odot$ to 3 $M_\odot$
%at 0.5 $M_\odot$ interval.
In Figure \ref{fig:Ldip}(a), we plot the results for selected models that are relevant for reflecting the main conclusions, which can be summarized as follows\footnote{Here we did not show any cases in which the NS collapses during spin-down process.}:
\begin{itemize}[leftmargin=*]
%\item For a given $P_0$, with the consideration of $R$ and $I$ evolution effects, the evolution history of magnetic dipole radiation luminosity $L_{\rm dip}$ would present three segments instead of two. Between $L_{\rm dip}\propto t^{0}$ and $L_{\rm dip}\propto t^{-2}$ segments, there emerges a new segment with
%$L_{\rm dip}\propto t^{-\gamma}$, where $\gamma>2$.

\item For the situations when realistic values are considered or the so called fiducial values are assigned to $R$ and $I$ (e.g. $R=10^6~\rm cm$ and $I=1.5\times10^{45}~\rm g~cm^2$), the magnetic dipole radiation luminosity
could be different up to one order of magnitude. In this case, if one use the so called fiducial values
to estimate the dipolar magnetic field strength and the initial spin period for the nascent NS, the results could
be systemically biased.

\item For a given EoS, different NS baryonic mass could evidently alter the magnitude and evolution history of
the dipole radiation luminosity. According to the results in Figure \ref{fig:Ldip}(a), with higher baryonic mass, the luminosity becomes smaller, and the spin-down timescale becomes larger. When the baryonic mass is large enough, the magnetar would collapse into a black hole after slightly spinning down, so that the dipole radiation would suddenly enter into a rapid descent phase.

\item For a given baryonic mass, different EoSs could overall change the magnitude of the dipole radiation luminosity. As shown in Figure \ref{fig:Ldip}(a), with a more stiff EoS, the luminosity generally becomes larger, and the spin-down timescale becomes smaller. Adopting different EoSs could change the fact that whether and when the magnetar would collapse.
\end{itemize}

Then we consider the GW radiation dominated case, where intensity of GW radiation depends on the NS asymmetries (manifested by the value of $\epsilon$).
Here we discuss two mechanisms to induce NS asymmetries: first, the crust of a NS is solid and elastic, where the solid shape is
related to the history of neutron star formation and EoS \citep{Haskell06,Lasky15}, in which scheme the $\epsilon$ value is usually
assumed to be a constant during the spin-down process \citep{Corsi09,Lasky16,Lu17,Sarin18,Lu19,Lan20,Sur21}. Second, the NS
asymmetry is caused by the distortion of NS magnetic field, in which scheme the ellipticity satisfies $\epsilon\propto B_{p}^{2}$
\citep{Bonazzola96,Haskell08,Dallosso09,Gao17a,Dallosso18,Lander20}. Similarly, two values of $M_b$ (e.g. $2.0M_{\odot}$ and $2.5M_{\odot}$) are tested. In figure \ref{fig:Ldip}(b) and Figure \ref{fig:Ldip}(c), we plot the results for both scenarios, which
can be summarized as follows:

\begin{itemize}[leftmargin=*]
\item  For both scenario, with realistic values of $R$ and $I$ for different EoSs and NS baryonic mass being considered, the magnetic dipole radiation luminosity could be variant up to one order of magnitude. Similar to the EM radiation dominated case, if one use the so called fiducial values of $R$ and $I$ to estimate properties of the nascent NS, the results would be significantly biased.
\item The history of magnetic dipole radiation luminosity for both scenario would consist four segments. The first two segments are $L_{\rm dip}\propto t^{0}$ followed by $L_{\rm dip}\propto t^{-\gamma}$. The last two segments would be $L_{\rm dip}\propto t^{-1}$ followed by $L_{\rm dip}\propto t^{-2}$. The emergence of a new segment $L_{\rm dip}\propto t^{-\gamma}$ is due to the evolution of $R$ and $I$. When $\epsilon$ is approximate to be a constant, $\gamma$ is slightly smaller than 1; when $\epsilon\propto B_{p}^{2}$, $\gamma$ is around or larger than 3.
\end{itemize}

In previous works, it has been proposed to obtain the
braking index of the nascent NS by fitting the magnetic dipole radiation luminosity with
\begin{eqnarray}
L_{\rm dip}=L_{\rm sd}(1+\frac{t}{T_{\rm sd}})^{\frac{4}{1-n}}
\label{L obs}
\end{eqnarray}
where $n$ is the braking index with $\dot{\Omega}\propto\Omega^{n}$, $L_{\rm sd}=B^2_{\rm p}R^{6}\Omega_0^{4}/6c^{3}$
is defined as the NS spin-down luminosity and $T_{\rm sd}\approx\frac{1}{2}I\Omega_0^{2}/L_{\rm sd}$ is defined
as the spin-down timescale of the NS. The obtained braking index is then used to diagnose whether the magnetar
spin-down is dominated by EM radiation or GW radiation \citep{Lasky17,Lu19}, since when EM
radiation dominates, the torque equation would become $\dot{\Omega}=-B_p^2R^6\Omega^3/6Ic^3$, so that the
braking index should be 3. On the other hand, when GW emission dominates, the torque equation would become
$\dot{\Omega}=-32GI\epsilon^2\Omega^5/5c^5$, so that the braking index would be 5. According to our results,
when the newly born NS has a millisecond spin period, $R$ and $I$ are no longer constant, so that if one still
apply Equation \ref{L obs} to fit the magnetic dipole radiation luminosity, the obtained braking index could
be slightly smaller than 3 for EM emission dominated case or greatly smaller than 3 for GW emission dominated
case with constant $\epsilon\propto B_{p}^{2}$, or larger than 5 for GW emission dominated case with constant
$\epsilon$.

On the other hand, our results show that the sharp decay $t^{-(>3)}$ following the X-ray plateaus \citep{Rowlinson10,Rowlinson13,Lu15,Sarin20a} are unnecessarily related to the collapsing of the central magnetar. Instead, it could be caused by the $R$ and $I$ evolution effect, if the spin-down process of the nascent NS is dominated by GW radiation and the ellipticity satisfies $\epsilon\propto B_{p}^{2}$. If our interpretation is correct, this may explain why some GRBs that present the internal plateau feature, still show late time central engine activity, manifested through flares and second shallow plateaus \citep{Troja07,Perley09,Margutti11,Gao15,Gao17b,Zhao20}. In the magnetar collapsing scenario, only flares or plateaus close to the internal plateau could be interpreted with fall-back accretion model \citep{chen17,Zhao20}.

\section{Analytical analysis for the numerical results}
Under certain approximations, we can try to explain the results from numerical calculation by analytical method. By fitting the evolutionary behavior of the $R$ and $I$ for different EoSs and baryonic mass, one can approximately get that
\begin{eqnarray}
\label{gp} R \simeq \left\{ \begin{array}{ll} R_0(\frac{\Omega}{\Omega_{\rm k}})^{m}, & \Omega_1<\Omega<\Omega_{\rm k};\\
R_0(\frac{\Omega_1}{\Omega_{\rm k}})^{m}, &
\Omega\leq\Omega_1. \\
\end{array} \right.
\end{eqnarray}

\begin{table*}
\begin{center}{\scriptsize
\caption{The evolution indexes of R and I with different EoSs and NS baryonic mass}
\begin{tabular}{ccccc}
\hline
\hline
\multicolumn{5}{c}{\qquad\qquad\qquad\qquad$m$}\\
\hline
& AP3 & ENG & SLy & WFF2 \\
\hline
$M_{b}=2.0M_{\odot}$ & 0.34 & 0.35 & 0.21 & 0.33 \\
$M_{b}=2.5M_{\odot}$ & 0.27 & 0.28 & 0.37 & 0.27 \\
$M_{b}=3.0M_{\odot}$ & 0.38 & NO & NO & NO \\
\hline
\multicolumn{5}{c}{\qquad\qquad\qquad\qquad$k$}\\
\hline
& AP3 & ENG & SLy & WFF2 \\
\hline
$M_{b}=2.0M_{\odot}$ & 0.14 & 0.13 & 0.13 & 0.12 \\
$M_{b}=2.5M_{\odot}$ & 0.15 & 0.24 & 0.25 & 0.22 \\
$M_{b}=3.0M_{\odot}$ & 0.31 & NO & NO & NO \\
  \hline
 \hline
 \end{tabular}
 }
\end{center}
\end{table*}

\begin{eqnarray}
\label{gp} I \simeq \left\{ \begin{array}{ll} I_0(\frac{\Omega}{\Omega_{\rm k}})^{k}, & \Omega_1<\Omega<\Omega_{\rm k};\\
I_0(\frac{\Omega_1}{\Omega_{\rm k}})^{k}, &
\Omega\leq\Omega_1, \\
\end{array} \right.
\end{eqnarray}
where $R_{0}$, $I_{0}$, and $\Omega_{\rm k}$ are the initial radius, initial moment of inertia, and Keplerian angular velocity of the newly born NS, respectively.
It is worth noting that $\Omega_{\rm k}$ depends on the NS EoS and baryonic mass. $m$ and $k$ are the power law index for $R$ and $I$ evolution with respect to the rotational speed of the NS. The best fitting values of $m$ and $k$ for different EoSs and $M_b$ are collected in Table 1. For our selected EoSs, $m$ ranges from 0.21 to 0.38 and $k$ ranges from 0.12 to 0.31. When the baryonic mass is large enough, the magnetar would collapse into a black hole after slightly spinning down, in which case the values of $m$ and $k$ are denoted as ``NO". With the assumption that the magnetic dipole moment $\mu\equiv B_pR^3$
is conserved, the evolution of $L_{\rm dip}$ could be derived based on the Equation \ref{dotE}:

(I) For $L_{\rm dip}$ dominated case, one has
\begin{eqnarray}
L_{\rm dip}\simeq -\frac{(2+k)I_0\Omega^{k+1}}{2\Omega_{\rm k}^{k}}\dot{\Omega}=\frac{B_{p,0}^2R_0^6\Omega^{4}}{6c^3},
\label{EM_dominated}
\end{eqnarray}
so that the complete solution of $\Omega(t)$ in Equation \ref{EM_dominated} could be written as
\begin{eqnarray}
\Omega=\Omega_0\left[1+\frac{t}{T_{\rm sd,em}}\right]^\frac{1}{k-2},
\label{Omega_EM}
\end{eqnarray}
where $\Omega_{0}$ is the initial angular frequency at $t=0$, and $T_{\rm sd,em}$ is a corresponding characteristic spin-down timescale, which could be derived as
\begin{eqnarray}
T_{\rm sd,em}=\frac{3(2+k)I_0c^3}{(2-k)B_{p,0}^2R_0^6\Omega_{\rm 0}^{2}}(\frac{\Omega_{\rm 0}}{\Omega_{\rm k}})^{k}.
\label{Tsdem}
\end{eqnarray}
Hence, the evolution history of $L_{\rm dip}$ could be expressed as
\begin{eqnarray}
L_{\rm dip}=L_{\rm sd,em}\left[1+\frac{t}{T_{\rm sd,em}}\right]^\frac{4}{k-2},
\label{Luminosity_EM}
\end{eqnarray}
where $L_{\rm sd,em}$ is the initial luminosity of electromagnetic dipole emission at $t=0$, which could be calculated as
\begin{eqnarray}
L_{\rm sd,em}=\frac{B_{p,0}^2R_0^6\Omega_0^{4}}{6c^3}.
\label{Lsdem}
\end{eqnarray}
It is clear that both the magnitude and the evolution history of $L_{\rm dip}$ depend on the realistic EoS and NS mass, as well as the evolution history of the moment of inertia (with index $k$). Comparing Equation \ref{L obs} with Equation \ref{Luminosity_EM}, one can obtain the braking index in this scenario as
\begin{eqnarray}
n=3-k.
\end{eqnarray}
When $k>0$, we have $n<3$.

(II) For $L_{\rm GW}$ dominated case with $\epsilon$ keeping constant, one has
\begin{eqnarray}
L_{\rm GW}\simeq -\frac{(2+k)I_0\Omega^{k+1}}{2\Omega_{\rm k}^{k}}\dot{\Omega}=\frac{32GI_0^2\epsilon^2\Omega^{6+2k}}{5c^5\Omega_{\rm k}^{2k}},
\label{GW_dominated}
\end{eqnarray}
so that the complete solution of $\Omega(t)$ in Equation \ref{GW_dominated} could be written as
\begin{eqnarray}
\Omega=\Omega_0\left[1+\frac{t}{T_{\rm sd,gw}}\right]^{-\frac{1}{k+4}},
\label{Omega_EM}
\end{eqnarray}
where $T_{\rm sd,gw}$ could be derived as
\begin{eqnarray}
T_{\rm sd,gw}=\frac{5(k+2)c^5}{64(k+4)GI_0\epsilon^2\Omega_0^{4}}(\frac{\Omega_{\rm 0}}{\Omega_{\rm k}})^{-k}.
\label{Omega evolution}
\end{eqnarray}
The evolution history of $L_{\rm dip}$ could be expressed as
\begin{eqnarray}
L_{\rm dip}=L_{\rm sd,em}\left[1+\frac{t}{T_{\rm sd,gw}}\right]^{-\frac{4}{k+4}}.
\label{Luminosity_GW}
\end{eqnarray}
Similar to the EM dominated case, here both the magnitude and the evolution history of $L_{\rm dip}$ depend on the realistic EoSs and NS mass, as well as the evolution history of the moment of inertia (with index $k$). Comparing Equation \ref{L obs} with Equation \ref{Luminosity_GW}, one can derive the braking index in this scenario as
\begin{eqnarray}
n=5+k.
\end{eqnarray}
When $k>0$, we have $n>5$.

(III) For $L_{\rm GW}$ dominated case with $\epsilon=\beta B_{p}^{2}$, one has
\begin{eqnarray}
L_{\rm GW}\simeq -\frac{(2+k)I_0\Omega^{k+1}}{2\Omega_{\rm k}^{k}}\dot{\Omega}=\frac{32GI_0^2\beta^2B^{4}_{p}\Omega^{6+2k}}{5c^5\Omega_{\rm k}^{2k}},
\label{GW2_dominated}
\end{eqnarray}
so that the complete solution of $\Omega(t)$ in Equation \ref{GW2_dominated} could be written as
\begin{eqnarray}
\Omega=\Omega_0\left[1+\frac{t}{T_{\rm sd,gw}}\right]^{\frac{1}{12m-k-4}},
\label{Omega_GW2}
\end{eqnarray}
where $T_{\rm sd,gw}$ could be derived as
\begin{eqnarray}
T_{\rm sd,gw}=\frac{5(k+2)R^{12}_{0}c^5}{64(k+4-12m)GI_0\beta^2\mu^{4}\Omega_0^{4}}(\frac{\Omega_{\rm 0}}{\Omega_{\rm k}})^{12m-k}.
\label{TGW evolution}
\end{eqnarray}
The evolution history of $L_{\rm dip}$ could be expressed as
\begin{eqnarray}
L_{\rm dip}=L_{\rm sd,em}\left[1+\frac{t}{T_{\rm sd,gw}}\right]^{\frac{4}{12m-k-4}}.
\label{Luminosity_GW2}
\end{eqnarray}
In this case, the evolution history of $L_{\rm dip}$ not only depends on the evolution history of the moment of inertia $I$, but also the evolution history of $R$. The decaying power law index $4/(12m-k-4)$ could be around or larger than 3. Comparing Equation \ref{L obs} with Equation \ref{Luminosity_GW2}, one can derive the braking index in this scenario as
\begin{eqnarray}
n=5+k-12m,
\end{eqnarray}
which could be much smaller than 5 (even smaller than 3).

\section{Conclusion and Discussion}

By solving the field equations, we find that when a NS's rotational speed is approaching the breakup limit, its radius and moment of inertia would undergo an obvious evolution as the NS spins down. In this case, the deceleration history of the NS would become more complicated for given initial dipole magnetic field and ellipticity. Our main results could be summarized as follows:

\begin{itemize}[leftmargin=*]
\item When realistic values of $R$ and $I$ for different EoSs and NS baryonic mass are considered, the magnetic dipole radiation luminosity could be variant within one to two orders of magnitude.

\item With the consideration of $R$ and $I$ evolution effects for a rapidly spinning NS, its magnetic dipole radiation light curve would present new segments. For instance,
when GW radiation dominates the spin down power and $\epsilon\propto B_{p}^{2}$, the history of magnetic dipole radiation luminosity would consist four segments, i.e., $L_{\rm dip}\propto t^{0}$ followed by $L_{\rm dip}\propto t^{-\gamma}$, and then followed by $L_{\rm dip}\propto t^{-1}$ and $L_{\rm dip}\propto t^{-2}$. The new segment $L_{\rm dip}\propto t^{-\gamma}$ with $\gamma$ larger than 3 is due to the evolution of $R$ and $I$. In this case, if one apply $L_{\rm dip}=L_{\rm sd}(1+t/T_{\rm sd})^{4/(1-n)}$ to fit the dipole radiation luminosity, the obtained braking index could be much smaller than 3 \footnote{Other reasons why the braking index could be smaller than 3 has been discussed in \cite{Lasky17}.}.

\item In the case when EM radiation power is comparable to the GW radiation power, the dipole radiation light curve would become even more complicated. Especially when the initial EM radiation power is slightly larger, for $\epsilon\propto B_{p}^{2}$ scenario, there would be a transition from EM loss dominating to GW loss dominating, and then back again as the NS spins down. In this case, the history of dipole radiation could consist five segments, i.e., $L_{\rm dip}\propto t^{0}$ followed by $L_{\rm dip}\propto t^{-2}$ and then followed by $L_{\rm dip}\propto t^{-\gamma}$ ($\gamma$ is larger than 3), and then followed by $L_{\rm dip}\propto t^{-1}$, finally followed by $L_{\rm dip}\propto t^{-2}$. For those complicated situations, Equation \ref{L obs} may not be a good formula to fit the dipole light curve. But if one still applies Equation \ref{L obs} to find the braking index, the result could vary from $n<3$ to $n>5$, depending on how many segments being covered by the observations.
\end{itemize}

According to our results, 1) we suggest that when using the EM observations such as the X-ray plateau data of GRBs to diagnose the properties of the nascent neutron stars, NS EoS and mass information should be invoked as simultaneously constrained parameters; 2) if the spin-down process of the nascent NS is dominated by GW radiation and the ellipticity satisfies $\epsilon\propto B_{p}^{2}$, the sharp decay following the X-ray plateau could be caused by the $R$ and $I$ evolution effect rather than due to the NS collapsing. If this is true, it may explain why some GRBs that present the internal plateau feature still show late time central engine activity, manifested through flares and second shallow plateaus. Future EM and GW joint detection could help to distinguish these two scenarios.

Finally, we would like to point out several caveats of our results. First, the physical conditions for a nascent NS would be very complicated, and some conditions may significantly alter the radiation lightcurve, so that the R/I evolution effect we discussed here would be reduced or even completely suppressed. For instance, it has been proposed that the evolution of the inclination angle between the rotation and magnetic axes of the NS could markedly revise the X-ray emission \citep{Cikintoglu20}. Moreover, the nascent NS is likely to undergo free precession in the early stages of its lifetime when the rotation and magnetic axes of the system are not orthogonal to each other, which would lead to the fluctuations in the X-ray light curve \citep{Suvorov20,Suvorov21}. On the other hand, the X-ray plateau emission luminosity could emerge from a plerion-like model of the nascent NS,
in which model the magnetized, relativistic wind from a millisecond magnetar injects shock-accelerated electrons into a cavity confined by
the GRB blastwave. The plerion model introduces an anticorrelation between the luminosity and duration of the plateau, and also shows a sudden
drop in the X-ray emission when the central magnetar collapses into a black hole \citep{Strang19}.

Secondly, even if the NS EoS and mass information were invoked as we suggest, one may still face many difficulties when using the X-ray plateau data of GRBs to diagnose the properties of the nascent NS. For instance, precise estimation for the luminosity of X-ray plateaus, in many cases, is very difficult to obtain, because there are no redshift measurement for most GRBs and the uncertainties introduced by the cosmological k-correction could also be significant \citep{Bloom01}. On the other hand, a constant efficiency is usually assumed to convert the spin-down luminosity to the observed X-ray luminosity, which might be incorrect. It has been proposed that the efficiency might be strongly dependent on the energy injecting luminosity, which leads to a larger braking index from afterglow fitting compared to the case with constant efficiency \citep{Xiao19}. Furthermore, if one considers both energy injection and radiative energy loss of the blastwave, the X-ray light curve would be altered without changing the braking index, and this model could also fit the data well \citep{Dallosso11,Sarin20b}.

Finally, it is worth noticing that besides the magnetar scenario, many alternative models have been proposed to interpret the X-ray plateau emission, such as the structure jet model \citep{Beniamini20}, the high-latitude emission model \citep{Oganesyan20}, and the late time energy injection model \citep{Matsumoto20}, etc. As suggestted by \cite{Sarin19}, before attempting to derive NS parameters, one should first ensure the magnetar scenario being the preferred hypothesis.

\acknowledgments
We thank the anonymous referee for the helpful comments that have helped us to improve the presentation of the paper. LL, HG and SZL acknowledge the National Natural Science Foundation of China under Grant No. 11722324, 11690024, 11633001, the Strategic Priority Research Program of the Chinese Academy of Sciences, Grant No. XDB23040100 and the Fundamental Research Funds for the Central Universities.

\end{document}